\documentclass[aps,pre,epsf,twocolumn,showpacs]{revtex4-1}
\usepackage{amsmath}
\usepackage{epsfig}
\usepackage{color}
\usepackage{epstopdf}
\usepackage{amssymb}
\usepackage{multirow}
\usepackage{hyperref}

\begin{document}

\title{Restoration of dimensional reduction in the random--field Ising model at five dimensions}

\author{Nikolaos G. Fytas$^{1}$, V\'{i}ctor Mart\'{i}n-Mayor$^{2,3}$, Marco Picco$^4$, and Nicolas Sourlas$^5$}

\affiliation{$^1$Applied Mathematics Research Centre, Coventry
University, Coventry CV1 5FB, United Kingdom}

\affiliation{$^2$Departamento de F\'isica T\'eorica I, Universidad
Complutense, 28040 Madrid, Spain}

\affiliation{$^3$Instituto de Biocomputac\'ion y F\'isica de
Sistemas Complejos (BIFI), 50009 Zaragoza, Spain}

\affiliation{$^4$Sorbonne Universit\'es, Universit\'e Pierre et
Marie Curie - Paris VI,
Laboratoire de Physique Th\'eorique et Hautes Energies, \\
4, Place Jussieu, 
75252 Paris Cedex 05, France}

\affiliation{$^5$Laboratoire de Physique Th\'eorique de l'Ecole
Normale Sup\'erieure (Unit{\'e} Mixte de Recherche du CNRS et de
l'Ecole Normale Sup\'erieure, associ\'ee \`a l'Universit\'e
Pierre et Marie Curie, PARIS VI)
  24 rue Lhomond, 75231 Paris CEDEX 05, France}

\date{\today}

\begin{abstract}
The random-field Ising model is one of the few disordered systems
where the perturbative renormalization group can be carried out to
all orders of perturbation theory. This analysis predicts
dimensional reduction, \textit{i.e.}, that the critical properties
of the random-field Ising model in $D$ dimensions are identical to
those of the pure Ising ferromagnet in $D-2$ dimensions. It is
well known that dimensional reduction is not true in three
dimensions, thus invalidating the perturbative renormalization
group prediction. Here, we report high-precision numerical
simulations of the 5D random-field Ising model at zero
temperature. We illustrate universality by comparing different
probability distributions for the random fields. We compute all
the relevant critical exponents (including the critical slowing
down exponent for the ground-state finding algorithm), as well as
several other renormalization-group invariants. The estimated
values of the critical exponents of the 5D random-field Ising
model are statistically compatible to those of the pure 3D Ising
ferromagnet. These results support the restoration of dimensional
reduction at $D = 5$. We thus conclude that the failure of the
perturbative renormalization group is a low-dimensional
phenomenon. We close our contribution by comparing universal
quantities for the random-field problem at dimensions $3 \leq D <
6$ to their values in the pure Ising model at $D-2$ dimensions and
we provide a clear verification of the Rushbrooke equality at all
studied dimensions.
\end{abstract}


\maketitle

\section{Introduction}
\label{sec:intro}

In the study of phase transitions under the presence of quenched
disorder~\cite{parisi:94}, the straightforward application of
field theoretic methods and the renormalization group (RG) is not
possible because the disorder breaks the translation symmetry of
the Hamiltonian. The standard procedure is then to average over
disorder using the replica method~\cite{edwards:75}. One starts
with $n$ noninteracting copies of the system (replicas) and
averages over the disorder distribution. This produces an
effective Hamiltonian with $n$ interacting fields which is
translation invariant and enables the use of the RG. In the end,
the $ n \to 0 $ limit has to be taken.

The replica method is mathematically unorthodox. Its combination
with the perturbative renormalization group (PRG) has been shown
to produce incorrect results in 3D systems. A warning example is
provided by the random-field Ising model (RFIM) where the
combination of the replica method with the PRG predicts
dimensional reduction~\cite{young:77,parisi:79c} (see below),
which does not hold neither in three~\cite{bricmont:87} nor in
four dimensions~\cite{fytas:16}. On the other hand, the replica
method has been proven correct in the case of branched polymers,
as well as for the highly non trivial problem of mean-field spin
glasses~\cite{talagrand:06,panchenko:13}. Mean field and the
replica method are believed to be correct at infinite dimensions.

The RFIM is probably the best studied problem in this context,
both for its simplicity and physical relevance. In fact, the RFIM
is one of the two well-known disordered systems (the other one
refers to the case of branched polymers) that can be analyzed to
all orders of perturbation theory, thanks to the existence of a
hidden supersymmetry~\cite{parisi:79c}. The PRG analysis predicts
the phenomenon of dimensional reduction: The critical properties
of the RFIM in $D$ dimensions should be the same as those of the
pure Ising ferromagnet at $D-2$ dimensions. It is by now well
established that this prediction is not true in three dimensions
because the 3D RFIM orders~\cite{bricmont:87}, while the 1D pure
Ising ferromagnet does not.

One central problem is to understand the reason of the failure of
the PRG. Since dimensional reduction is proven to all orders of
perturbation theory, the reasons of its failure must be non
perturbative. Parisi and Sourlas argue that in the case of the
RFIM in three dimensions the interaction between replicas is
attractive and leads to the formation of bound states between
replicas~\cite{parisi:02}. The presence of bound states is a non
perturbative phenomenon. The mass of the bound state provides a
new length scale which is not taken into account in the
traditional PRG analysis. The authors of Ref.~\cite{parisi:02}
also provide a physical interpretation: These bound states
indicate that the correlation length is not
self-averaging~\cite{parisi:02}.

\begin{figure}
\centerline{\includegraphics[scale=.30, angle=0]{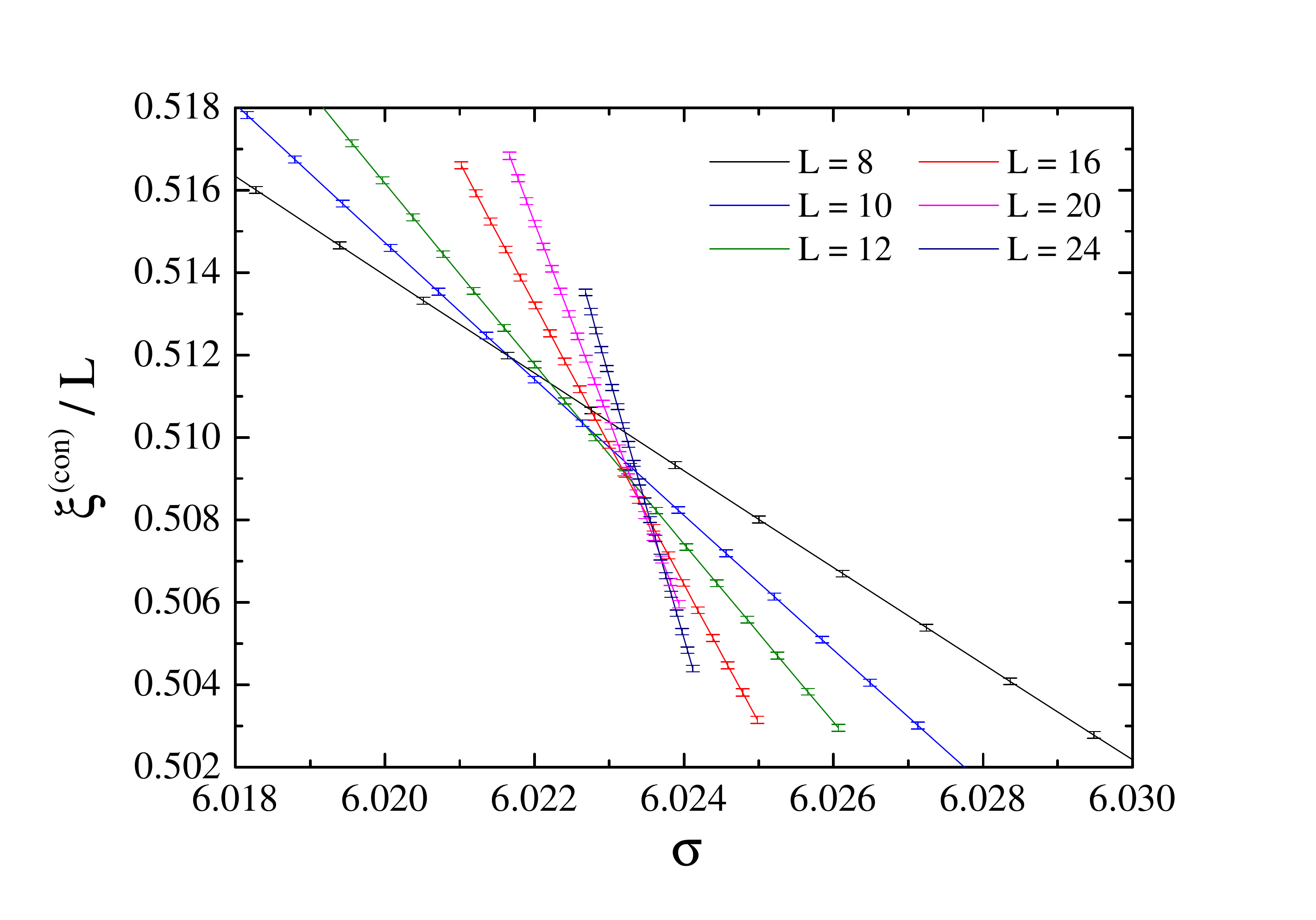}}
\caption{Connected correlation length in units of the system size
$L$ vs. $\sigma$ for the Gaussian 5D RFIM (we show data only for
some characteristic $L$ values for clarity' sake). Due to scale
invariance, all curves should cross at the critical point
$\sigma_\mathrm{c}$. Yet, small systems deviate from the large-$L$
scale-invariant behavior.} \label{fig:quotients}
\end{figure}

Although the finding of Parisi and Sourlas is numerical, we have
more indications for the presence of bound states. Br\'ezin and De
Dominicis have also noticed that the forces between replicas are
attractive and that the Bethe-Salpeter kernel, for a pair of
replicas of different indices, develops an instability for $ D \le
6$ hinting towards the existence of bound states among
replicas~\cite{brezin:98,brezin:01}. Similar conclusions were
reached by Kardar and coworkers, who studied the problem of 2D
interfaces~\cite{kardar:87,kardar:87b,kardar:89,kardar:93}. Indeed, using
the Bethe ansatz method, these authors solved the replica
Hamiltonian, thus finding that bound states form when the number
of replicas is $n < 1$. Bound states were also found in the case
of the random Potts ferromagnet in two
dimensions~\cite{parisi:04}.

Identifying the existence (or lack thereof) of bound states as the
crucial factor for the validity of the PRG immediately suggests
that the space dimension should play a crucial role. In fact, we
know from constructive field theory that in the formation of bound
states there is a competition between the attractive interactions
and the size of the available phase space. In two dimensions the
phase space is small and any infinitesimal attraction is enough to
form bound states~\cite{dimock:76}. The size of the phase space
increases when the dimension of space gets larger. In higher
dimensions the formation of bound states depends on the strength
of the attractive forces. We expect that for high enough
dimensions bound states will no longer exist, thus implying that
the PRG prediction of dimensional reduction should eventually
hold.

\begin{figure}
\centerline{\includegraphics[scale=.50, angle=0]{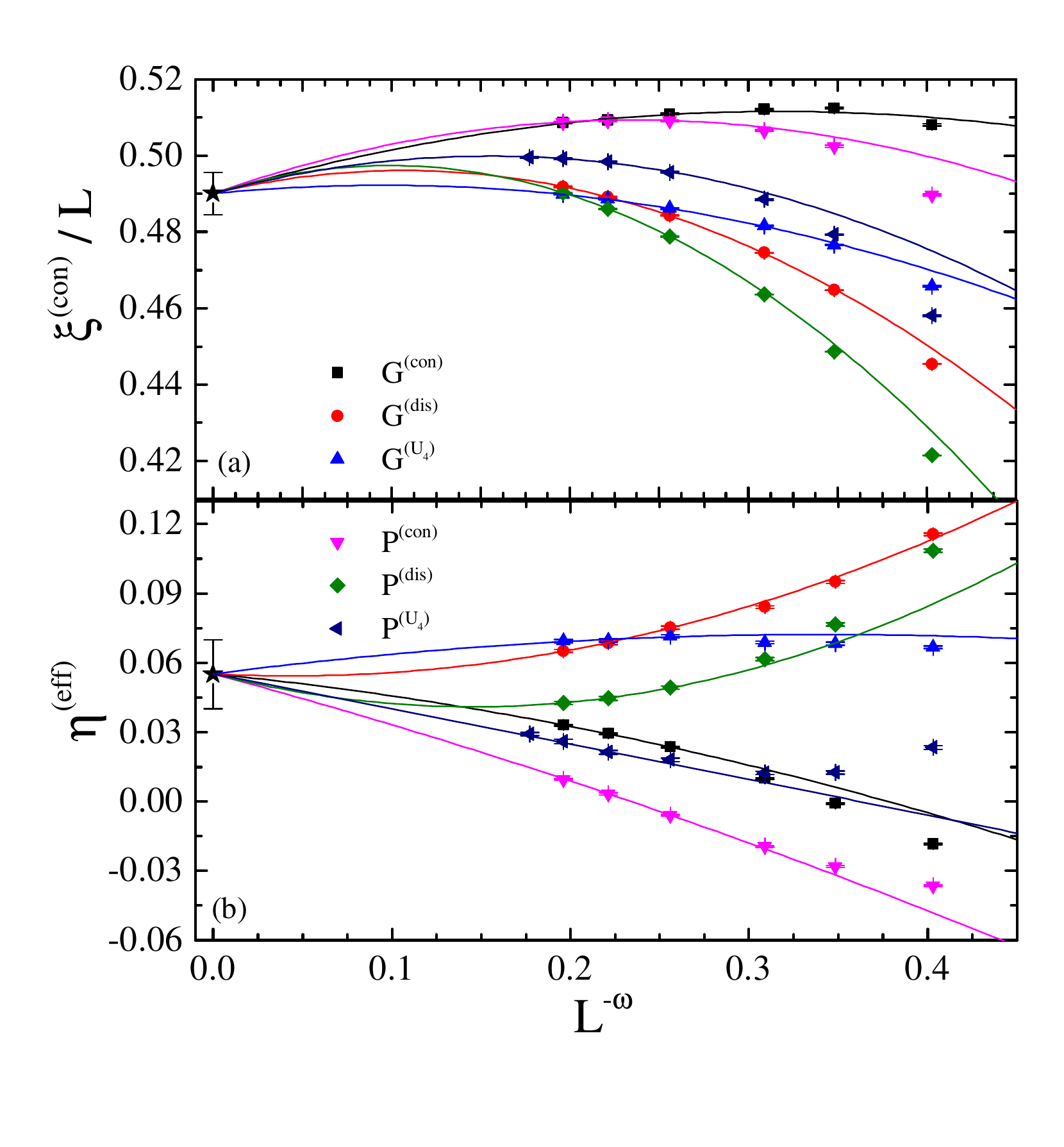}}
\caption{$\xi^{\mathrm{(con)}}/L$ (top) and
$\eta^{\mathrm{(eff)}}$ (bottom) vs. $L^{-\omega}$. Lines
correspond to the joint quadratic ($a_3 = 0$) fit~(\ref{QF})
reported in Tab.~\ref{tab:quotients} with $\omega=0.66$. The
points at $L^{-\omega} = 0$ mark our infinite volume extrapolation
with their error bars.} \label{fig:omega}
\end{figure}

The natural question to ask then is if there exists an
intermediate dimension $D_{\mathrm{int}}$ below the upper critical
dimension $D_{\rm u} = 6$~\cite{aharony:78} such that the PRG and
replicas are valid for dimensions $D > D_{\mathrm{int}}$ and false
for $D < D_{\mathrm{int}}$. This $D_{\mathrm{int}}$ may depend on
the physical system. Tarjus \emph{et al.} using functional RG
arguments concluded that such $D_{\mathrm{int}}$ exists for the
RFIM and that it is close to $D=5$. In particular they found
$D_{\mathrm{int}} \simeq
5.1$~\cite{tissier:11,tissier:12,tarjus:13}.

Here we report large-scale zero-temperature numerical simulations
of the RFIM at five spatial dimensions. Our analysis benefits from
recent advances in finite-size scaling and reweighting methods for
disordered systems~\cite{fytas:13,fytas:15b}. By using two
different random-field distribution we are able to show the
universality of the critical exponents characterizing the
transition. Our results are compatible with dimensional reduction
being restored in five dimensions: We find that the critical
exponents of the 5D RFIM are compatible to those of the pure 3D
Ising ferromagnet up to our relatively small simulation errors,
and in agreement to the suggestion by Tarjus \emph{et
al.}~\cite{tissier:11,tissier:12,tarjus:13}.

The outline of the article is as follows: In Sec.~\ref{sec:model}
the model and methods employed are described shortly and in
Sec.~\ref{sec:results} our main results on the universality
principle and the critical exponents of the 5D RFIM are presented.
We conclude this article in Sec.~\ref{sec:summary} by providing an
overview of the model's critical behavior in dimensions $3\leq D <
D_{\rm u}$, which is compared to that of the pure Ising
ferromagnet under the prism of dimensional reduction.

\begin{figure}
\centerline{\includegraphics[scale=.30, angle=0]{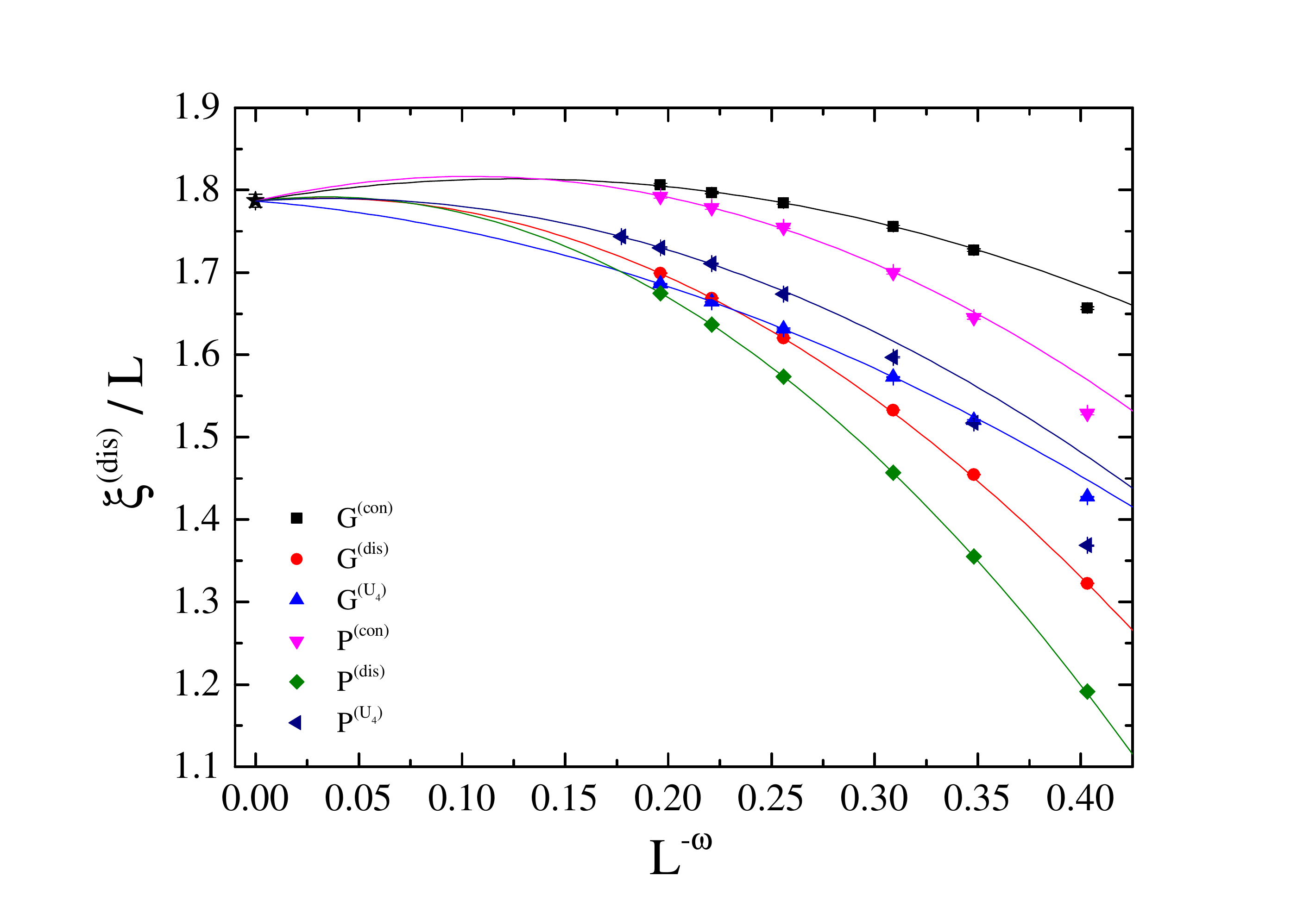}}
\caption{$\xi^{\mathrm{(dis)}}/L$ vs. $L^{-\omega}$. Lines
correspond to a joint quadratic ($a_3 = 0$) fit~(\ref{QF}) with
$\omega=0.66$. The point at $L^{-\omega} = 0$ marks our infinite
volume extrapolation with its error bar.} \label{fig:xi_dis}
\end{figure}

\section{Model and Methods}
\label{sec:model}

The Hamiltonian of the RFIM is
\begin{equation}
\label{H} {\cal H} = - J \sum_{<xy>} S_x S_y - \sum_{x} h_x S_x \;
,
\end{equation}
with the spins $S_x = \pm 1$ on a hypercubic lattice in $D$
dimensions with nearest-neighbor ferromagnetic interactions and
$h_x$ independent random magnetic fields with zero mean and
variance $\sigma$. A given realization of the random fields
$\{h_x\}$ is named a sample. Because the disorder is quenched, one
first takes thermal mean values for a sample, denoted as
$\langle\cdots \rangle$, and only then average over samples, which
we indicate by an over-line (for instance, for the magnetization
density $m=\sum_x S_x/L^D$ we consider first $\langle m\rangle$
and then $\overline{\langle m \rangle}$).

It is established that the relevant fixed point of the model lies
at zero temperature~\cite{villain:84,bray:85,fisher:86}.
Therefore, the only spin configuration that we shall consider in
the present work is the ground state for each specific realization
of the Hamiltonian~(\ref{H}) on a $D=5$ hypercubic lattice with
periodic boundary conditions and energy units $J=1$. Our random
fields $h_{x}$ follow either a Gaussian $({\mathcal P}_G)$, or a
Poissonian $({\mathcal P}_P)$ distribution:
\begin{equation}
{\mathcal P}_G(h,\sigma) = {1\over \sqrt{2 \pi
      \sigma^2}} e^{- {h^2 \over 2\sigma^2}}\;,\ {\mathcal P}_P(h,\sigma) = {1\over 2 |\sigma| } e^{- {|h| \over
\sigma}}\;,
\end{equation}
where $-\infty< h < \infty$. As it is clear, for both
distributions $\sigma$ is our single control parameter.

There are two relevant propagators for the RFIM, namely the
connected, $C^{\mathrm{(con)}}$, and disconnected one
$C^{\mathrm{(dis)}}$. At the critical point and for large $r$ ($r$
being the distance between $x$ and $y$) they decay as:
\begin{eqnarray}\label{eq:definition-of-C-connected}
C^{\mathrm{(con)}}_{xy}&\equiv&\frac{\partial\overline{\langle
  S_x\rangle}}{\partial h_y} \ \sim\
\frac{1}{r^{D-2+\eta}}\,,\\\label{eq:definition-of-C-disconnected}
C^{\mathrm{(dis)}}_{xy}&\equiv&
\overline{\langle S_x\rangle\langle S_y\rangle}\ \sim\
\frac{1}{r^{D-4+\bar\eta}}\,.
\end{eqnarray}
The above expressions define as well the two relevant anomalous
dimensions, $\eta$ and $\bar\eta$. For each of these two
propagators we shall consider the second-moment correlation
lengths~\cite{amit:05}, denoted as $\xi^{\mathrm{(con)}}$ and
$\xi^{\mathrm{(dis)}}$, respectively. Hereafter, we shall indicate
with the superscript ``$\mathrm{con}$'', \emph{e.g.}
$\xi^{\mathrm{(con)}}$, quantities computed from the connected
propagator. Similarly, the superscript ``$\mathrm{dis}$'',
\emph{e.g.} $\xi^{\mathrm{(dis)}}$, will refer to the  propagator
$C^{\mathrm{(dis)}}$.

\begin{figure}
\centerline{\includegraphics[scale=.30, angle=0]{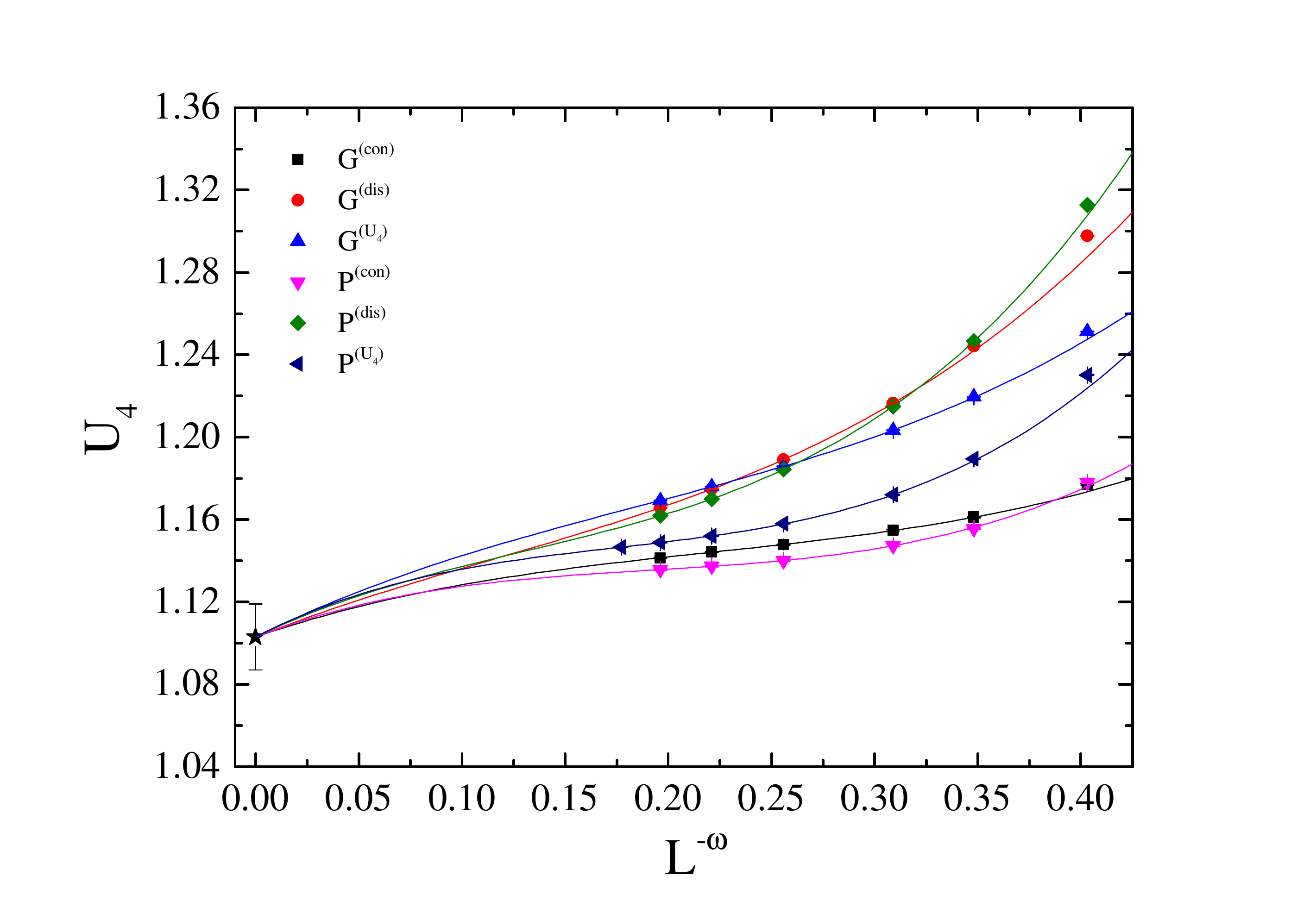}}
\caption{$U_{4}$ vs. $L^{-\omega}$. Lines correspond to a joint
cubic fit~(\ref{QF}) with $\omega=0.66$. The point at $L^{-\omega}
= 0$ marks our infinite volume extrapolation with its error bar.}
\label{fig:U4}
\end{figure}

We simulated lattice sizes from $L_{\rm min}=4$ to $L_{\rm
max}=28$. For each $L$ and $\sigma$ value we computed ground
states for $10^7$ samples. For comparison: $5000$ samples of
$L_{\rm max} = 14$ were simulated in Ref.~\cite{ahrens:11}. Our
simulations and analysis closely follow our previous work at $D=3$
and $4$~\cite{fytas:13,fytas:16} (see Ref.~\cite{fytas:15b} for
full details). Thus, we just briefly recall here the fundamental
aspects of our computation.

The algorithm used to generate the ground states of the system was
the push-relabel algorithm of Tarjan and
Goldberg~\cite{goldberg:88}. We prepared our own C version of the
algorithm, involving a modification proposed by Middleton \emph{et
al.}~\cite{middleton:01,middleton:02,middleton:02b} that removes
the source and sink nodes, reducing memory usage and also
clarifying the physical
connection~\cite{middleton:02,middleton:02b}. Additionally, the
computational efficiency of our algorithm has been increased via
the use of periodic global
updates~\cite{middleton:02,middleton:02b}.

From simulations at a given $\sigma$, we computed $\sigma$-derivatives
and extrapolated to neighboring $\sigma$ values by means of a
reweighting method~\cite{fytas:13,fytas:15b}.

We also computed the corresponding
susceptibilities $\chi^{\mathrm{(con)}}$ and
$\chi^{\mathrm{(dis)}}$, as well as the dimensionless Binder ratio
$U_4 = \overline{\langle m ^4 \rangle}/\overline{\langle
m^2\rangle}^2$ and the ratio
$U_{22}=\chi^{\mathrm{(dis)}}/[\chi^{\mathrm{(con)}}]^2$ that
gives a direct access to the difference of the anomalous
dimensions $2\eta-\bar\eta$~\cite{fytas:13,fytas:15b}.

We followed the quotients-method approach to finite-size
scaling~\cite{amit:05,nightingale:76,ballesteros:96}. In this
method one considers dimensionless quantities $g(\sigma,L)$ that,
barring correction to scaling, are $L$-independent at the critical
point. We consider three such $g$, namely
$\xi^{\mathrm{(con)}}/L$, $\xi^{\mathrm{(dis)}}/L$, and $U_4$.
Given a dimensionless quantity $g$, we consider a pair of lattices
sizes $L$ and $2L$ and determine the crossing
$\sigma_{\mathrm{c},L}$, where $g(\sigma_{\mathrm{c},L},L)=
g(\sigma_{\mathrm{c},L},2L)$, see Fig.~\ref{fig:quotients}. For
each random-field distribution we computed three such
$\sigma_{\mathrm{c},L}$, a first for $\xi^{\mathrm{(con)}}/L$,
another for $\xi^{\mathrm{(dis)}}/L$, and a third for $U_4$.
Crossings approach the critical point $\sigma_{\mathrm{c}}$ as
$\sigma_{\mathrm{c}}-\sigma_{\mathrm{c},L}={\cal
  O}(L^{-(\omega+1/\nu)})$, with $\omega$ being the leading
corrections-to-scaling exponent.

Dimensionful quantities $O$ scale with $\xi$ in the thermodynamic
limit as $\xi^{x_O/\nu}$, where $x_O$ is the scaling dimension of
$O$. At finite $L$, we consider the quotient $Q_{O,L} =
O_{2L}/O_L$ at the crossing (for dimensionless magnitudes $g$, we
write $g^\mathrm{cross}_{L}$ for either $g_{L}$ or $g_{2L}$,
whichever shows less finite-size corrections)
\begin{equation}\label{eq:QO}
Q_{O,L}^\mathrm{cross} = 2^{ x_O/\nu} +
  O(L^{-\omega}) \; \; ; \; \; g^\mathrm{cross}_{L} = g^{\ast} +
  O(L^{-\omega})\,.
\end{equation}
$Q_{O}^\mathrm{cross}$ (or $g^\mathrm{cross}_{L}$) can be
evaluated at the crossings of $\xi^{\mathrm{(con)}}/L$,
$\xi^{\mathrm{(dis)}}/L$, and $U_4$. The three choices differ only
in the scaling corrections, an opportunity we shall use. The RG
tells us that $x_O$, $g^\ast$, $\omega$, and $\nu$, are universal.
We shall compute the critical exponents using Eq.~\eqref{eq:QO}
with the following dimensionful quantities: $\sigma$-derivatives
[$x_{D_\sigma
    \xi^{\mathrm{(con)}}}= x_{D_\sigma \xi^{\mathrm{(dis)}}}=1+\nu$],
susceptibilities [$x_{\chi^{\mathrm{(con)}}}= \nu(2-\eta)$ and
  $x_{\chi^{\mathrm{(dis)}}}= \nu(4-\bar\eta)$] and the ratio $U_{22}$
[$x_{U_{22}}=\nu(2\eta-\bar\eta)$].

\begin{figure}
\centerline{\includegraphics[scale=.30, angle=0]{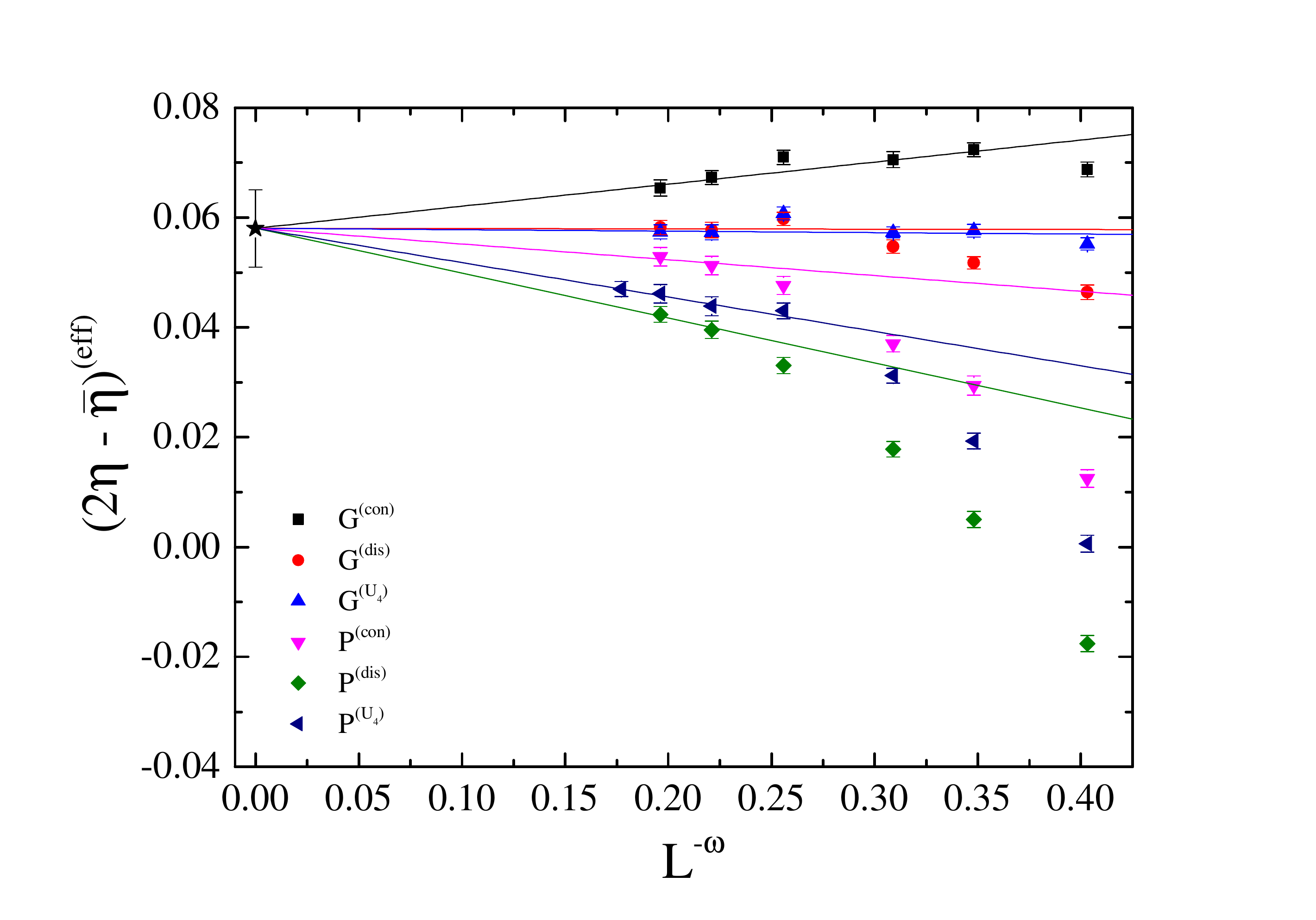}}
\caption{Effective anomalous dimension difference $(2\eta
-\bar\eta)^{\rm (eff)}$ vs. $L^{-\omega}$. Lines correspond to a
joint leading-term ($a_2 = a_3 = 0$) fit~\eqref{QF} with
$\omega=0.66$. The point at $L^{-\omega} = 0$ marks our infinite
volume extrapolation with its error bar.} \label{fig:2eta-bareta}
\end{figure}

As we applied the quotients method at the crossings of $\xi^{\rm
(con)} / L$, $\xi^{\rm (dis)} / L$, and $U_4$, typically the data
sets of our simulations were tripled for each pair of system sizes
used. Note also, that throughout the manuscript we shall use the
notation $\rm {Z}^{\rm (x)}$, where Z denotes the distribution - G
for Gaussian and P for Poissonian - and the superscript x the
crossing type considered - (con), (dis), or ($U_4$) - for
$\xi^{\rm (con)} / L$, $\xi^{\rm (dis)} / L$, and $U_4$,
respectively.

Since the size evolution can be non-monotonic as will be also seen
below in the relevant figures, and given that our accuracy is
enough to resolve sub-leading corrections to scaling, we take
these into account in an effective way: Let $X_L$ be either
$g^\mathrm{cross}_{L}$ or the effective scaling dimension
$x^\mathrm{(eff)}_O/\nu = \log Q_{O}^\mathrm{cross}(L)/\log 2$,
recall Eq.~\eqref{eq:QO}. We consider the following generalized
fitting functions
\begin{eqnarray}
X_L&=& X^{\ast}+a_1 L^{-\omega}+ a_2 L^{-2\omega} + a_3 L^{-3\omega}\; ,\label{QF}\\
\sigma_{\mathrm{c},L}&=&\sigma_c+ b_1 L^{-(\omega+\frac{1}{\nu})}+
b_2 L^{-(2\omega+\frac{1}{\nu})} \;,\label{QFS}
\end{eqnarray}
where $a_k$, with $k=1,2,3$, and $b_l$, with $l=1,2$, are scaling
amplitudes.

For the fitting procedure discussed below we restricted ourselves
to data with $L\geq L_\mathrm{min}$. As usual, to determine an
acceptable $L_\mathrm{min}$ we employed the standard $\chi^2$-test
for goodness of fit, where $\chi^2$ was computed using the
complete covariance matrix. Specifically, the $p$-value of our
$\chi^2$-test -- also known as $Q$, see \emph{e.g.}
Ref.~\cite{press:92} -- is the probability of finding a $\chi^2$
value which is even larger than the one actually found from our
data. Recall that this probability is computed by assuming: (i)
Gaussian statistics and (ii) the correctness of the fit's
functional form. We consider a fit as being fair only if $10\% < Q
< 90\%$. Generally speaking, we observed that, once a fair fit is
found, increasing $L_\mathrm{min}$ doubles (or worsens) the errors
in the extrapolation to $L=\infty$. However, increasing the order
of $L^{-\omega}$ in fits to Eq.~\eqref{QF} is even more
detrimental to the error in the extrapolation $X^*$. Therefore, we
first decide the order of the fit. Starting from linear in
$L^{-\omega}$ corrections to scaling, we increase $L_\mathrm{min}$
from $L_\mathrm{min}=4$ and check if the resulting fit is
acceptable (\emph{i.e.}, whether or not the $p$-value satisfies
our constraint $10\% < Q < 90\%$). If the fit's quality is not
acceptable, we increase $L_\mathrm{min}$ to the larger available
$L$ and try to fit again. In the case where we exhaust the number
of available system sizes without finding a fair fit, we move on
to quadratic scaling corrections. If an $L_\mathrm{min}$ yielding
an acceptable fit cannot be identified, then we consider
corrections to scaling of order $L^{-3\omega}$. As a rule, we keep
the lowest order for which an acceptable $L_\mathrm{min}$ can be
found. Having decided the order of $L^{-\omega}$ in
Eq.~\eqref{QF}, we also keep the smallest possible
$L_\mathrm{min}$.

\section{Evidence for Dimensional Reduction at $D = 5$}
\label{sec:results}

The procedure we follow is standard by now~\cite{ballesteros:98b}.
The first step is the estimation of the corrections-to-scaling
exponent $\omega$. Take, for instance, $\xi^{\mathrm{(con)}}/L$.
For each pair of sizes $(L,2L)$ we have six estimators: Three
crossing points, $\xi^{\mathrm{(con)}}/L$,
$\xi^{\mathrm{(dis)}}/L$, and $U_4$, and two disorder
distributions, Gaussian and Poissonian. Rather than six
independent polynomial fits to some degree of Eq.~\eqref{QF}, we
perform a single joint fit: We minimize the combined $\chi^2$
goodness-of-fit, by imposing that the extrapolation to $L=\infty$
(depicted as a black star at the origin of the horizontal axis for
all the figures below), $(\xi^{\mathrm{(con)}}/L)^{\ast}$, as well
as exponent $\omega$ are common for all six estimators (only the
scaling amplitudes differ). We judge from the final $\chi^2$ value
whether or not the fit is fair.

\begin{figure}
\centerline{\includegraphics[scale=.30, angle=0]{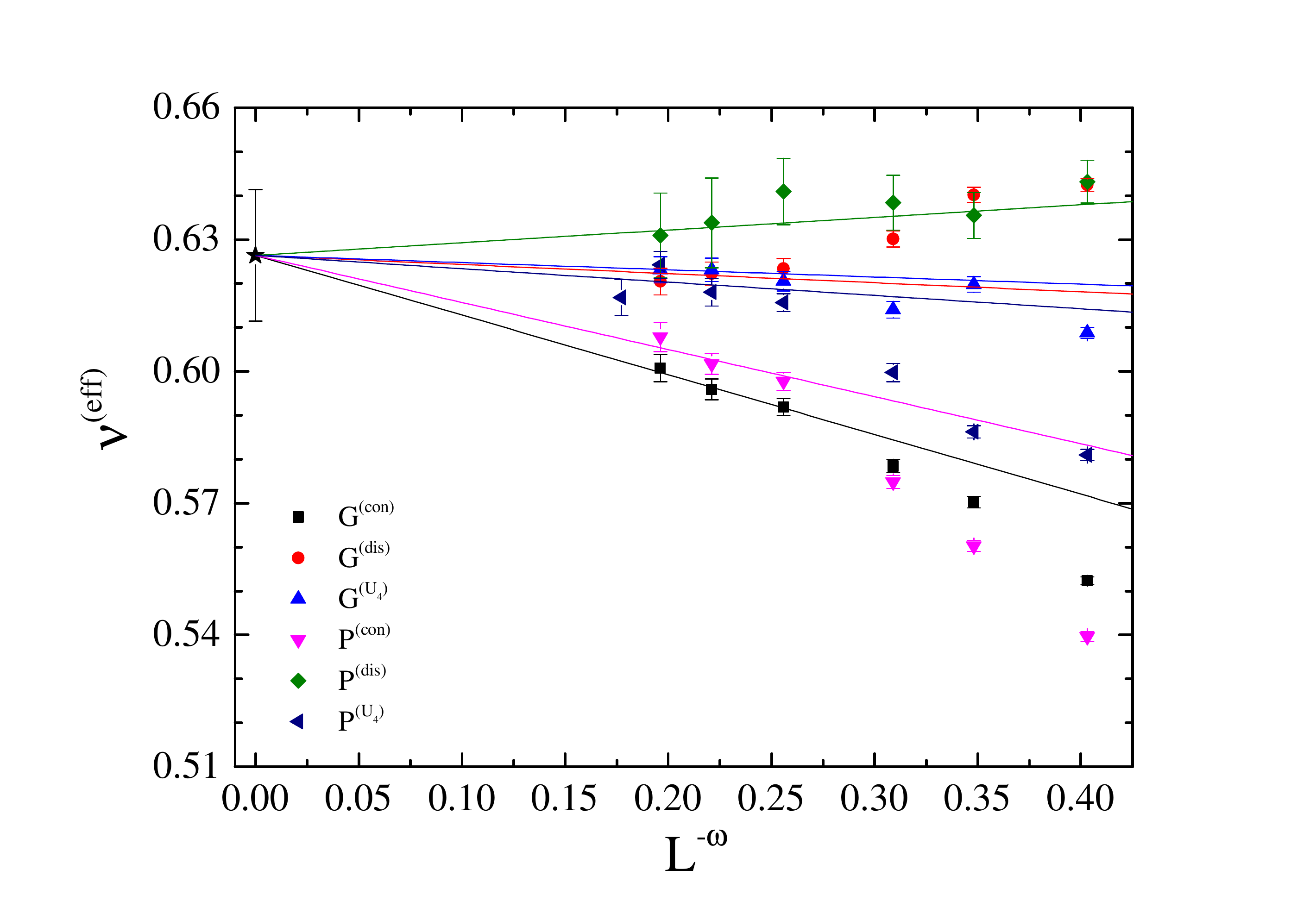}}
\caption{Effective critical exponent $\nu^{\rm (eff)}$ vs.
$L^{-\omega}$. Lines correspond to a joint leading-term ($a_2 =
a_3 = 0$) fit~(\ref{QF}) with $\omega=0.66$. The point at
$L^{-\omega} = 0$ marks our infinite volume extrapolation with its
error bar.} \label{fig:nu}
\end{figure}

\begin{figure}
\centerline{\includegraphics[scale=.30,
angle=0]{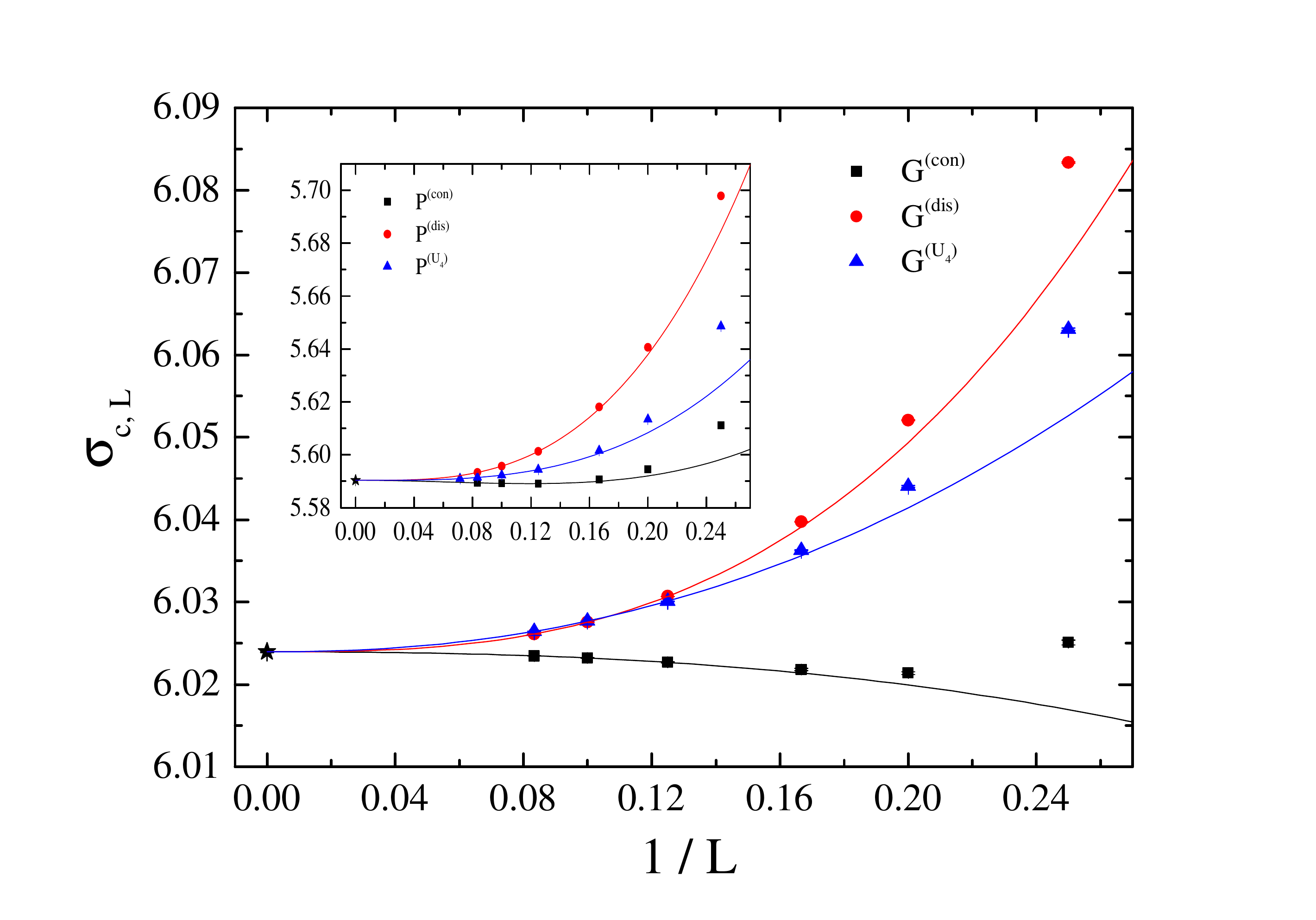}} \caption{Crossing points
$\sigma_{\mathrm{c},L}$ for Gaussian (main panel) and Poissonian
(inset) random fields. Lines are fits to Eq.~\eqref{QFS},
constrained to yield a common extrapolation at $L = \infty$.}
\label{fig:critical_fields}
\end{figure}

Furthermore, one can perform joint fits for several magnitudes,
say $\xi^{\mathrm{(con)}}/L$ and $\eta$. Of course, the
extrapolation to $L=\infty$ is different for each magnitude, but a
common $\omega$ is imposed. However, when we increase the number
of magnitudes, the covariance matrix becomes close to singular due
to data correlation and the fit unstable. Therefore, we limit
ourselves to $\xi^{\mathrm{(con)}}/L$ and $\eta$, see
Fig.~\ref{fig:omega}. We obtain a fair fit by considering pairs
$(L,2L)$ with $L\geq L_{\mathrm{min}}=8$, see
Tab.~\ref{tab:quotients}. Indeed there are not many other
available choices of pairs of observables to be considered in a
joint fit, unless one is willing to consider third-order
corrections to scaling (see for instance the data for $U_4$ in
Fig.~\ref{fig:U4}). Given that our lattice sizes range from $L=4$
up to $L=28$, we prefer to keep the order of the scaling
corrections as low as possible in the computation of $\omega$.

The rest of the quantities of interest are individually
extrapolated, following the same procedure, but now fixing
$\omega=0.66$, the value obtained in the joint fit of
Fig.~\ref{fig:omega}. For the extrapolation of the dimensionless
quantities $\xi^{\mathrm{(dis)}}/L$ and $U_4$ we refer the reader
to Figs.~\ref{fig:xi_dis} and \ref{fig:U4}. The extrapolation of
the difference $2\eta-\bar\eta$ and the critical exponent $\nu$ of
the correlation length are illustrated in
Figs.~\ref{fig:2eta-bareta} and \ref{fig:nu}, respectively. In
particular, in Fig.~\ref{fig:2eta-bareta} we show $\log
U_{22}/\log 2$ which is a direct measurement of the difference
$2\eta -\bar\eta$ and in Fig.~\ref{fig:nu} the effective values of
$\nu$ estimated as the derivatives of $\xi^{\rm (dis)}$ for all
data sets at hand (the statistical errors of the other
$\nu$-estimators were rather large and therefore omitted from the
fits). Finally, in Fig.~\ref{fig:critical_fields} the critical
fields for both the Gaussian (main panel) and Poissonian (inset)
5D RFIM are estimated via a joint fit of the form~\eqref{QFS}.

The final values we quote for all our observables are summarized
in Tab.~\ref{tab:quotients}. In fact, the extrapolations in
Tab.~\ref{tab:quotients} have two error bars. The first error,
obtained from the corresponding joint fit to Eqs.~\eqref{QF} and
\eqref{QFS} is of statistical origin. The second error is
systematic and takes into account how much the extrapolation to
$L=\infty$ changes within the range $0.53 < \omega < 0.81$.

\begin{table}
\caption{\label{tab:quotients} Summary of results for the 5D RFIM.
The first column is the outcome of a fit to Eq.~\eqref{QF}
(critical points $\sigma_\mathrm{c}$ were obtained from
Eq.~\eqref{QFS}, respectively). The second column is the standard
figure of merit,
 $\chi^2/{\rm DOF}$, where DOF denotes the number of
degrees of freedom. The third column gives the minimum system size
used in the fits and the last column the degree of the polynomial
in $L^{-\omega}$. The first set of rows reports a joint fit for
$\xi^{\mathrm{(con)}}/L$, $\eta$, and $\omega$. The remaining
quantities were individually extrapolated to $L=\infty$. The error
induced by the uncertainty in $\omega$ is given as a second error
estimate in the square brackets.}
\begin{ruledtabular}
\begin{tabular}{r@{\,=\,}lccc}
\multicolumn{2}{c}{Extrapolation to $L \rightarrow \infty$} & $\chi^{2}/{\rm DOF}$ & $L_{\rm min}$ & order in $L^{-\omega}$\\
\hline
$\xi^{\rm (con)}/L$&$0.4901(55)$               & \multirow{3}{*}{$11.3/10$}     & \multirow{3}{*}{8}  & \multirow{3}{*}{second}     \\
$\eta$&$0.055(15)$        &     &     &       \\
$\omega$&$0.66(+15/-13)$                   &     &     &       \\
\hline $\xi^{\rm (dis)}/L$&$1.787(8)[+30/-82] $ & $5.3/9$ & 6 &
second       \\
$U_4$&$1.103(16)[+18/-43] $ & $1.9/6$ & 6 &
third     \\
$2\eta-\bar{\eta}$&$0.058(7)[+1/-2] $      & $3.8/6$ & 10  &
first     \\
$\nu$&$0.626(15)[+2/-3] $      & $8.3/6$ & 10 &
first     \\
$\sigma_{\rm c}({\rm G})$&$6.02395(7)[+2/-7] $ &
$0.1/2$      &   8  & second    \\
$\sigma_{\rm c}({\rm P})$&$5.59038(16)[+9/-13]$               & $2.7/3$      &   8  &  second    \\
\end{tabular}
\end{ruledtabular}
\end{table}

\begin{figure}
\centerline{\includegraphics[scale=.30, angle=0]{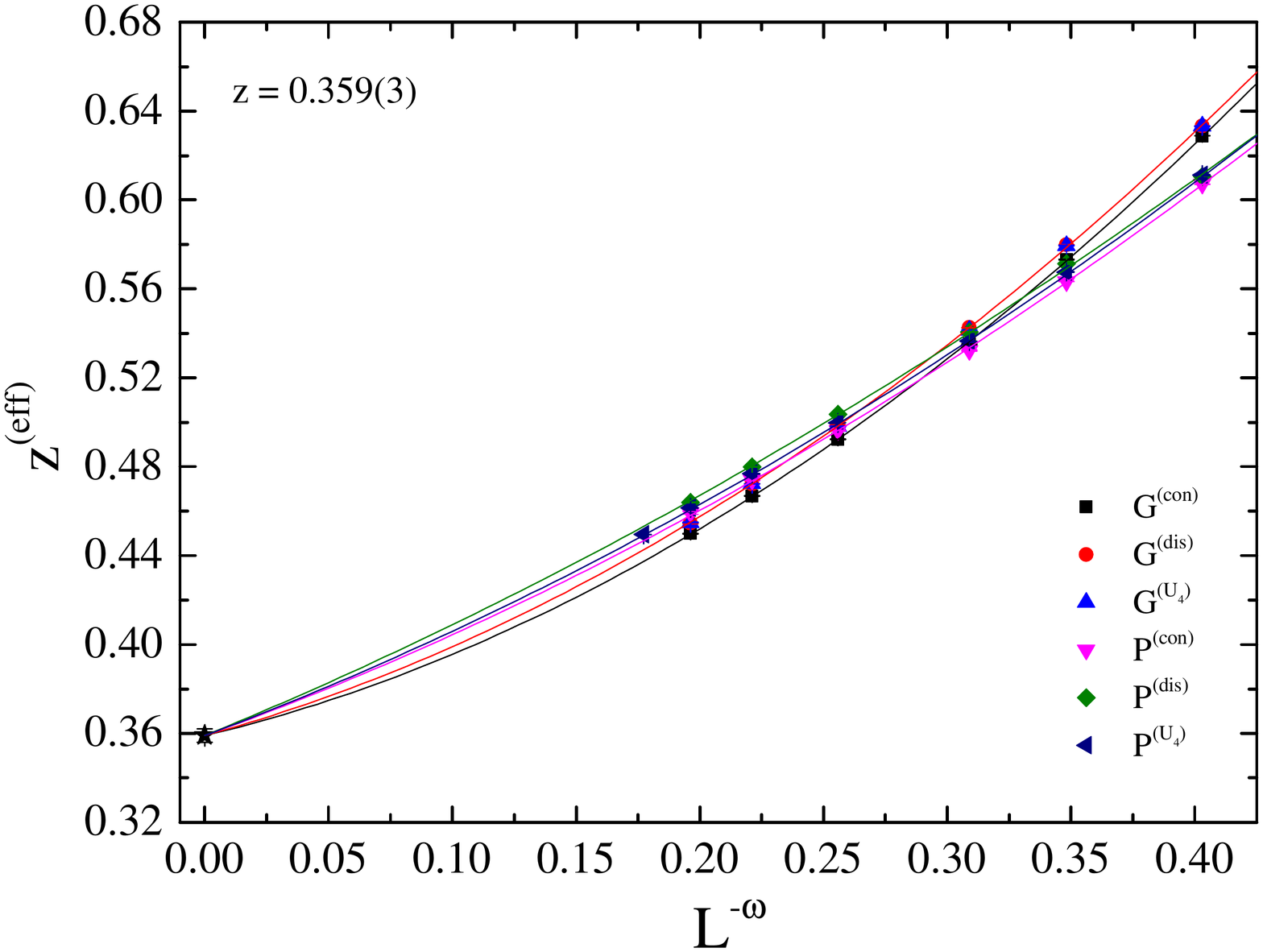}}
\caption{Effective critical slowing down exponent $z^{\rm (eff)}$
of the push-relabel algorithm vs. $L^{-\omega}$. The point at
$L^{-\omega} = 0$ marks our infinite volume extrapolation with its
error bar.} \label{fig:exponent_z}
\end{figure}

\begin{table*}
\caption{\label{tab:null_hypothesis} Fitting tests and results
after accepting the null hypothesis of restoration of dimensional
reduction at $D = 5$. The first two columns refer to the
observable and the $L\rightarrow \infty$ extrapolation,
respectively. The third column gives the standard figure of merit
$\chi^{2}/{\rm DOF}$. The fourth column is the $p$-value of our
$\chi^2$ tests (namely, the probability of $\chi^2$ to be even
larger than what we actually found, should the null hypothesis
hold). Finally, the fifth and sixth columns give the minimum size
used in the fits and the degree of the polynomial in
$L^{-\omega}$. The first two row-sets of results refer to the
joint fit~\eqref{QF}, by fixing in the first case $\omega$ to the
Ising value and in the second case both $\omega$ and the
extrapolated value of $\eta$ to their Ising values. The following
rows refer to either universal ratios, critical exponents, or
critical points. For the case of universal ratios we have fixed
$\omega$ to the Ising value, whereas for the cases of
$2\eta-\bar{\eta}$ and $\nu$ we ave fixed both $\omega$ to the
Ising value but also their extrapolation values to
$2\eta-\bar{\eta} = \eta = 0.036298$ and $\nu = 0.629971$.
Finally, for the case of critical points we have fixed both
$\omega$ and $\nu$ to the Ising values in the fits. The values of
$\omega$, $\eta$, and $\nu$ of the 3D Ising universality class
have been taken from Ref.~\cite{kos:16}.}
\begin{ruledtabular}
\begin{tabular}{llcccc}
Observable & Extrapolation to $L \rightarrow \infty$ & $\chi^{2}/{\rm DOF}$ & $p$-value & $L_{\rm min}$ & order in $L^{-\omega}$\\
\hline $\xi^{\rm (con)}/L$ & $0.4972(+16/-35)$  &    &    &  &     \\
$\eta$&$0.0453(+19/-44)$        &  $13.37/11$   &  $27\%$ & 8 &   second    \\
$\omega$&$0.82966$ (fixed)              &     &     &     &       \\
\hline
$\eta$&$0.036298$  (fixed)      &  $15.82/12$   & $20\%$  &8  &  second     \\
$\omega$&$0.82966$ (fixed)              &     &     &     &       \\
\hline \hline $\xi^{\rm (dis)}/L$&$1.8184(52)
 $ & $13.08/9$ & $16\%$ & 6 &
second       \\
$U_4$&$1.123(8) $ & $2.76/6$ & $84\%$ & 6 &
third     \\
 $2\eta-\bar{\eta}$&$0.036298$  (fixed)      & $4.15/7$ & $76\%$ & 8
& second     \\
 $\nu$&$0.629971 $ (fixed)     & $3.43/7$ & $84\%$ &
8 &
second     \\
 $\sigma_{\rm c}({\rm G})$&$6.02393(18)
 $ &
$0.95/2$    & $62\%$ &   8  & second    \\
$\sigma_{\rm c}({\rm P})$&$5.59028(13)
$               & $2.01/3$ &  $57\%$   &   8  &  second    \\
\end{tabular}
\end{ruledtabular}
\end{table*}

At this point several comments are in order:

\begin{itemize}

\item For dimensionless quantities we needed a second-order
polynomial in $L^{-\omega}$ to extrapolate our data, apart from $U_4$ where a cubic term
was necessary for the fit. On the other hand, leading-order
corrections sufficed for a safe estimation of the critical
exponent $\nu$ and the difference $2\eta - \bar\eta$.

\item We are not aware of any other previous computation of the
corrections-to-scaling exponent $\omega$ and of the dimensionless
ratios $\xi^\mathrm{(con)}/L$, $\xi^\mathrm{(dis)}/L$, and $U_4$
in the 5D RFIM. As it was shown above in Figs.~\ref{fig:omega},
\ref{fig:xi_dis}, and \ref{fig:U4}, all of them are universal and
together with the recently reported results of the 3D and 4D
RFIM~\cite{fytas:13,fytas:16}, they provide a complete picture of
universality in terms of different field distributions in the
random-field problem.

\item Our values for the critical exponents $\eta$ and $\nu$
(including the corrections-to-scaling exponent $\omega$ and the
difference $2\eta-\bar\eta$) are compatible within statistical
accuracy to the values of the pure 3D Ising ferromagnet: $\eta
=2\eta - \bar\eta = \bar\eta = 0.036298(2)$,  $\nu = 0.629971(4)$,
and $\omega = 0.82966(9)$~\cite{kos:16}, thus indicating that
within simulation errors, dimensional reduction gets restored at
five dimensions. As it can be seen from the results of
Tab.~\ref{tab:quotients}, a larger deviation among the computed
exponents and those of the Ising universality appears in the
anomalous dimensions. Of course, the computation of such small
numbers is a harsh task.

\item Notwithstanding, one would like to have a clear-cut answer
to the following important question: \emph{Are the critical
exponents of the 5D RFIM (even to our high accuracy) compatible to
those of the 3D pure Ising ferromagnet?} In order to answer
quantitatively the question, we make the null-hypothesis of
equality of the two universality classes. Indeed, in
Tab.~\ref{tab:null_hypothesis} we provide the figure of merit
$\chi^2/\mathrm{DOF}$, as well as the corresponding $p$-values,
for fits where the extrapolation to $L=\infty$ and the
corrections-to-scaling $\omega$ were taken from the 3D pure Ising
universality class. All fits, for which the extrapolation to
$L=\infty$ is known for the 3D pure Ising ferromagnet, are denoted
as (fixed) in Tab.~\ref{tab:null_hypothesis}. We remark that those
3D extrapolations are known to such a high-accuracy~\cite{kos:16},
that we can regard them as virtually exact. As the reader can
check in Tab.~\ref{tab:null_hypothesis}, for all the
extrapolations assuming 3D pure Ising universality we could
identify an appropriate $L_\mathrm{min}$ that makes the fit fair.
So, the answer to the above question is that at least within our
level of accuracy (which is set by the results shown in
Tab.~\ref{tab:quotients}), the two universality classes of the 5D
RFIM and the 3D pure Ising ferromagnet, cannot be distinguished.

\item We note the discrepancy in the determination of the critical
point for the Gaussian RFIM: Ref.~\cite{ahrens:11} quotes
$\sigma_{\rm c}(G) = 6.0157(10)$. This difference is  probably
explained by the fact that in Ref.~\cite{ahrens:11}
corrections-to-scaling were not taken into account and that our
statistics is much higher.

\end{itemize}

\begin{table*}
\caption{Illustrative summary of results for the $D$-dimensional
RFIM, where $D = 3$, $4$, and $D = 5$. In particular four row-sets
of results are shown: critical exponents (first set), the
verification of the Rushbrooke relation (second set), critical
points (third set), and universal ratios and the
corrections-to-scaling exponent $\omega$ (fourth set). For the
case of the critical exponent $\alpha$ we show two estimates, one
direct estimation~\cite{fytas:15b,fytas:16b} and another one based
on the modified hyperscaling relation. Corresponding results of
the 2D and 3D pure Ising ferromagnet are also included in the
fifth and sixth columns for comparison. The last column contains
mean-field (MF) results.} \label{tab:summary}
\begin{center}
\begin{tabular}{ | l || c | c  | c | c | c | c| }
\hline
            &     3D RFIM~\cite{fytas:13,fytas:15b}     & 4D RFIM~\cite{fytas:16,fytas:16b}         & 5D RFIM (current work)           & 2D IM~\cite{mccoy:73} & 3D IM~\cite{kos:16} & MF \\
\hline
$\nu$   &    1.38(10)  & 0.8718(58) &  0.626(15)   &  1 &   0.629971 (4)  &   1/2    \\
$\eta$   &  0.5153(9) & 0.1930(13) &  0.055(15)   &  0.25 & 0.036298(2)              &    0   \\
$\bar\eta$& 1.028(2)& 0.3538(35) &   0.052(30)   &   0.25 & 0.036298(2)            &    0   \\
$\Delta_{\eta,\bar\eta} = 2 \eta -\bar\eta$ &0.0026(9)& 0.0322(23) & 0.058(7)  &  0.25 & 0.036298(2)       &     0  \\
$\beta$  & 0.019(4)  & 0.154(2) & 0.329(12) & 0.125 & 0.326419(3) & 1/2 \\
$\gamma$ & 2.05(15) & 1.575(11)& 1.217(31) & 1.875 & 1.237075(10) & 1 \\
$\theta$    & 1.487(1) & 1.839(3)& 2.00(2) & 2 & 2 & 2 \\
$\alpha$   &  -0.16(35)  & 0.12(1) & - & - & - & - \\
$\alpha$ (from hyperscaling)  &  -0.09(15) & 0.12(1) & 0.12(5)  & 0 & 0.110087 (12) & 0 \\
 \hline\hline
$\alpha + 2 \beta + \gamma $  &  2.00(31) & 2.00(3) & 2.00(11)  & 2 & 2.000000 (28) & 2 \\
 \hline\hline
$\sigma_{\mathrm{c}} (G)$   & 2.27205(18) & 4.17749(6) & 6.02395(7)   & - & - & -  \\
$\sigma_{\mathrm{c}} (P)$  & 1.7583(2) & 3.62052(11) & 5.59038(16)  & - & - & -  \\
\hline \hline
$U_4$    & 1.0011(18) & 1.04471(46) & 1.103(16)    & &   &  \\
$\xi^{\mathrm{(con)}}/L$   & 1.90(12)   & 0.6584(8) & 0.4901(55)   & &   & \\
$\xi^{\mathrm{(dis)}} / L $  & 8.4(8)  & 2.4276(70) & 1.787(8) & & &   \\
$\omega $ & 0.52(11) & 1.30 (9) & 0.66(+15/-13) &   & 0.82966(9)  &  0 \\
\hline
\end{tabular}
\end{center}
\end{table*}

Finally, we discuss some computational aspects of the implemented
push-relabel algorithm and its performance on the study of the
RFIM. Although its generic implementation has a polynomial time
bound, its actual performance depends on the order in which
operations are performed and which heuristics are used to maintain
auxiliary fields for the algorithm. Even within this polynomial
time bound, there is a power-law critical slowing down of the
push-relabel algorithm at the zero-temperature
transition~\cite{ogielski:85}. A direct way to measure the
dynamics of the algorithm is to examine the dependence of the
running time, measured by the number of push-relabel operations,
on system size $L$~\cite{middleton:01,middleton:02,middleton:02b}.
Such an analysis has been carried out for the 3D and 4D versions
of the model and a FIFO (first in, first out) queue
implementation~\cite{middleton:01,middleton:02,middleton:02b,fytas:15b,fytas:16b}.
We present here results for the performance of the algorithm on
the 5D RFIM using our scaling approach within the quotients method
and numerical data for both Gaussian and Poissonian random-field
distributions. In Fig.~\ref{fig:exponent_z} we plot the effective
exponent values of $z$ at the various crossing points considered,
as indicated in the panel. The solid line is a joint quadratic
($a_3 = 0$) fit of the form~(\ref{QF}) with $\omega=0.66$. The
obtained estimate for the dynamic critical exponent is $z =
0.359(3)$, as marked by the filled star at $L^{-\omega} = 0$.

\section{Summary of results for the RFIM at $3\leq D < 6$}
\label{sec:summary}

We find it most useful to present in this last Section a summary
of the most recent computations of the critical properties of the
RFIM at three and higher dimensions by our
group~\cite{fytas:13,fytas:15b,fytas:16,fytas:16b}, still below
the upper critical dimensionality $D_{\rm u} = 6$. Our
presentation will take place under the prism of the original
prediction of dimensional reduction, by contrasting the critical
exponents of the $D$-dimensional RFIM to those of the pure $D-2$
Ising ferromagnet. The current numerical data at hand will also
allow us to further verify some of the most controversial scaling
relations in the literature of the RFIM, that is the Rushbrooke
relation $\alpha + 2 \beta + \gamma = 2$. In doing so, we will
make use of some standard exponent relations to provide estimates
for the complete spectrum of critical exponents.

In particular:

\begin{itemize}

\item The violation of the hyperscaling exponent $\theta$ may be
estimated via the anomalous dimensions $\eta$ and $\bar\eta$ as
$\theta = 2 -\bar\eta+\eta = 2-\eta+\Delta_{\eta,\bar\eta}$, where
$\Delta_{\eta,\bar\eta} = 2\eta-\bar\eta$.

\item We note here also the relation of $\theta$ to the critical
exponent $\alpha$ of the specific heat via the modified
hyperscaling relation $(D - \theta)\nu = 2 - \alpha$, which then
leads to $\alpha=2-\nu(D- 2 +\eta -\Delta_{\eta,\bar{\eta}})$.

\item Finally, for the estimation of the magnetic critical
exponents $\beta$ and $\gamma$ we have used the standard relations
$\beta = \nu(D - 4 + \bar\eta)/2 $ and $\gamma = \nu (2-\eta)$.

\end{itemize}

In Tab.~\ref{tab:summary} we present all our results for the
critical exponents, critical points, and universal ratios of the
RFIM at $D = 3$, $4$, and $D = 5$. The first and most striking
observation is that the critical exponents of the 4D RFIM have a
clear deviation when compared to those of the 2D Ising ferromagnet
indicating the breaking of dimensional reduction at the
dimensionality $D = 4$ and pointing at $D_{\rm int} > 4$, as has
already been stressed in Ref.~\cite{fytas:16}. On the other hand,
the deviation from the supersymmetry ($\eta = \bar\eta$ or $\theta
= 2$) clearly decreases with increasing $D$ and our numerical
results at five dimensions are compatible, within statistical
accuracy, to a restoration of the supersymmetry at $D = 5$ (see
also the statistical tests presented in
Tab.~\ref{tab:null_hypothesis} that support our claim). As
discussed above, the measured exponents of the 5D RFIM are close
to those of the pure 3D Ising ferromagnet, but not exactly the
same when it comes to the anomalous dimensions. This still leaves
open the possibility that the restoration takes places at a
(non-physical) real value of $D$ slightly larger than $5$ and not
exactly at $D = 5$. Another important remark of
Tab.~\ref{tab:summary} is that our numerical estimates for the
critical exponents $\alpha$, $\beta$, and $\gamma$, satisfy the
Rushbrooke relation up to a very high accuracy and at all studied
dimensions $D = 3$, $4$, and $D = 5$.

So, where do we stand at this point? Clearly, we have now at hand
a complete picture of the model's critical behavior for $D <
D_{\rm u}$. This includes very accurate estimates of the full
spectrum of critical exponents, critical points, and universal
ratios, as well as an unarguable claim of universality and the
verification of scaling relations. These latter concepts have been
severely questioned in the study of the random-field problem but
now seem to be perfectly settled. What may be seen as a further
step in the study of the random-field problem would be a detailed
investigation of criticality at the suspected upper critical
dimension $D_{\rm u} = 6$, for which characteristic logarithmic
scaling violations have been reported~\cite{ahrens:11}, but still
await for a detailed confirmation.

To conclude, let us point out that the questions addressed in this
paper are of interest for the properties of phase transitions of
disordered systems in general, and not only for the RFIM. Still,
the RFIM is unique among other models due to the existence of very
fast algorithms that make the study of these questions numerically
feasible.

\acknowledgments{We thank the staff of the BIFI supercomputing center for their
assistance and for computing time at the cluster \emph{Memento}. This work was
partially supported by MINECO (Spain) through Grant No. FIS2015-65078-C2-1-P.
N.~G.~F. and M.~P. were supported by a Royal Society International Exchanges
Scheme 2016/R1. N.~G.~F. is grateful to Coventry University for providing
a Research Sabbatical Fellowship during which part of this work has been completed.}

\bibliographystyle{apsrev4-1}

\bibliography{biblio}

\end{document}